\documentclass[11pt,disablejfam]{article}
\usepackage{fancybox}
\usepackage{amsmath,amssymb}
\usepackage{color}

\newcommand{\hh}{\hspace{2mm}}
\newcommand{\h}{\hspace{1mm}}
\newcommand{\bz}{\bar{z}}

\begin{document}

\begin{titlepage}

\vspace*{-10ex}
\begin{flushright}
OU-HET 630\\
KEK-TH-1316\\
June 7,  2009
\end{flushright}
\bigskip

\begin{center}
{\LARGE \bf  Left-Right Asymmetric Holographic RG  Flow
\\
\vskip2mm
with Gravitational Chern-Simons Term
}
\vspace{15ex}

\setcounter{footnote}{0}

{\renewcommand{\thefootnote}{\fnsymbol{footnote}}}
{\large\bf 
Kyosuke Hotta$^{a}$
\footnote{E-mail: {\tt hotta@het.phys.sci.osaka-u.ac.jp}}, 
Yoshifumi Hyakutake$^{b}$
\footnote{E-mail: {\tt hyaku@post.kek.jp}}, 
Takahiro Kubota$^{a}$
\footnote{E-mail: {\tt kubota@cep.osaka-u.ac.jp}},
\\ 
Takahiro Nishinaka$^{a}$
\footnote{E-mail: {\tt nishinaka@het.phys.sci.osaka-u.ac.jp}}
and 
Hiroaki Tanida$^{a}$
\footnote{E-mail: {\tt hiroaki@het.phys.sci.osaka-u.ac.jp}}}

\vspace{5ex}

$^{a}$
{\sl Department of Physics, Graduate School of Science, Osaka
 University,
\\
Toyonaka, Osaka 560-0043, Japan
\\
$^{b}$High Energy Accelerator Research Organization (KEK), \\
Tsukuba, Ibaraki, 305-0801, Japan
}
\end{center}

\vspace{10ex}

\begin{abstract}
We consider the holographic renormalization group (RG) flow 
in three dimensional gravity with the gravitational Chern-Simons term  
coupled to some scalar fields.  We apply the canonical approach 
to this higher derivative case and employ the Hamilton-Jacobi formalism
to analyze the flow equations of two dimensional field theory. 
Especially we obtain flow equations of Weyl and gravitational anomalies,
and derive $c$-functions for left and right moving modes.
Both of them are monotonically non-increasing along the flow, and the difference 
between them is determined by the coefficient of the gravitational Chern-Simons term. 
This is completely consistent with the Zamolodchikov's 
$c$-theorem for parity-violating  two-dimensional quantum field theories. 
\end{abstract}

\end{titlepage}

\pagestyle{plain}


\section{Introduction}

The three dimensional Einstein gravity has been an intriguing 
theoretical laboratory to study classical as well as quantum gravity. 
When the negative cosmological constant is added, the theory exhibits 
even more interesting properties. 
It has the three dimensional anti-de Sitter 
space (AdS$_3$) as a vacuum \cite{DJ}, and a black hole solution can be constructed 
out of the AdS$_3$ geometry by an  appropriate identification \cite{btz}. 
Brown and Henneaux \cite{brown} succeeded to uncover  the conformal 
symmetry on the boundary of the AdS$_3$ space 
and derived the left and right Virasoro algebras which share the same central 
charge. 
 The Cardy's asymptotic formula of 
the state counting with the help of 
the  central charge reproduces 
 the correct thermodynamical entropy formula \cite{strominger98}.

In \cite{HHKT}, we examined  the canonical 
formalism of gravity when  the gravitational Chern-Simons term 
is added to the Einstein-Hilbert action  with the negative cosmological 
constant. Such a theory is often referred to as topologically 
massive gravity (TMG) \cite{DJT} and received much attention in 
recent years \cite{chiralgravity} - \cite{anninos}. 
We investigated the conformal symmetry living on the boundary of the AdS$_3$ 
space and showed, by the canonical method, that the left and right Virasoro 
central charges are shifted by an amount equal in magnitude but of 
opposite sign (see also \cite{krauslarsen} - 
\cite{guptasen}).

In our previous paper \cite{HHKNT}, 
we studied the three dimensional Einstein gravity without the 
Chern-Simons term, while a scalar field was  added, 
paying attention to the gauge/gravity correspondence 
\cite{Maldacena} - \cite{Witten}.
Given a certain type of scalar potential, we presented a 
black hole   solution, whose  metric  
exhibits a peculiar property that the spacetime geometry is AdS$_{3}$ 
both at spatial infinity and at the horizon. 
The Virasoro algebras and their central charges were obtained 
by applying the  Brown-Henneaux's method both at the spatial 
infinity and at the horizon
\footnote{At the horizon we employed covariant formulation
of charges \cite{BB,BC}, which is paid much attention recently to study 
the Kerr/CFT correspondence \cite{ghss}.}. 
Regarding the radial direction in the bulk as 
the renormalization scale of the dual boundary theory, we  discussed 
renormalization group (RG) flow \cite{BVV} - \cite{berg}
and have defined the $c$-function \cite{Zamo} that 
is monotonically non-increasing toward the infrared direction.

In the present paper we discuss in further detail the RG flow,   
by including the gravitational Chern-Simons term into the 
three dimensional Einstein gravity coupled to scalar fields. 
In  the presence of the gravitational Chern-Simons term, the action is not strictly invariant under the 
diffeomorphism, but receives variations from 
the boundary term. According to the AdS/CFT correspondence, the non-invariance 
under the diffeomorphism in the bulk manifests itself in the non-conservation 
of the boundary energy-momentum tensor, i.e., the gravitational anomaly 
\cite{krauslarsen}. 
Such an anomaly effect was closely connected to the asymmetric central 
charges of the left and right moving Virasoro algebras.

We apply the canonical formalism adapted for the case of higher derivative 
theories \cite{Ostrogradsky,BL,BLK,HHKT} 
and derive the Hamilton-Jacobi equation. 
Thereby we  discuss the RG flow of the boundary field theory 
regarding the radial coordinate as the energy scale. 
By looking at the Weyl anomaly of energy momentum tensor of the boundary theory, we read off  the sum 
of the left and right $c$-functions.
We notice that the energy momentum tensor itself is not defined in a covariant way.
However, the Bardeen-Zumino polynomial naturally arises in the Hamilton-Jacobi equation,
and makes a modified energy momentum tensor transforming covariantly.

We  will go one step further to analyze the momentum constraint 
to obtain the gravitational anomaly relation on the boundary, 
and identify the difference of the two $c$-functions 
from its coefficient.
The left and right $c$-functions thus obtained  are monotonically 
non-increasing  toward the infrared region  and settle down 
to the central charges of the boundary CFT. 
As it turns out, the difference between the left and right 
$c$-functions is independent of the renormalization scale. 
This is shown to be consistent with the 
derivation of the left-right asymmetric $c$-functions for 
parity-violating two-dimensional quantum field theories \cite{chiral-c}.

The organization of this paper is as follows.
In section 2, we briefly review $c$-functions of the parity-violating two dimensional quantum field theory.
In section 3, we analyze holographic RG flow of the three dimensional gravity 
with gravitational Chern-Simons term, and derive the left-right asymmetric $c$-functions.
Section 4 is devoted to conclusion and discussion.
An explicit form of the solution in the gravity theory is given in appendix A.


\section{
The left-right asymmetric $c$-functions and anomalies
}

In this section, we review the 
 $c$-theorem \cite{Zamo} for 
parity-violating two-dimensional quantum field theory
and define $c$-functions for left- and right-movers, 
$c_{L}(t)$ and $c_{R}(t)$, which  monotonically decrease along 
the renormalization group flow \cite{chiral-c}. 
We emphasize  the fact that the difference of the two functions is 
constant along the RG flow, namely 
\begin{eqnarray}
t\frac{d}{dt}\left (c_L (t)- c_R (t)\right )=0,
\label{eq.1}
\end{eqnarray} 
where $t$ is the scaling parameter to be defined below.
At the fixed point, if we consider the curved two-dimensional
background, the difference of two central charges can be read off from 
 the gravitational anomaly, while the sum of two central charges can
 be related to the Weyl anomaly. We will briefly discuss these points 
  later in this section. For now, however, 
  we consider flat two-dimensional spacetime to keep close contact with 
  the original proof of  Zamolodchikov's $c$-theorem.




By using complex coordinates $z$ and $\bz$, the conservation law and 
the symmetrical property of the energy momentum tensor can be written as
\begin{eqnarray}
 \partial \bar{T} + \bar{\partial}\Theta = 0,\quad \bar{\partial}T +
  \partial\Theta = 0, \label{T-conserv}
\end{eqnarray}
where we define $\partial = \partial/\partial z,\, \bar{\partial} =
\partial/\partial\bar{z},\, T=T_{zz},\, \bar{T}=T_{\bar{z}\bar{z}}$ and $\Theta =
T_{z\bar{z}} = T_{\bar{z}z}$. We use these equations to study the
scaling properties of two point functions of the energy-momentum tensor.
For example, note that by using the right equation
we can exchange $\bar{\partial} T$ for
$-\partial\Theta$. Then we find
\begin{eqnarray}
 \bar{\partial}\left<T(z,\bar{z})T(0,0)\right> = -
  \partial\left<\Theta(z,\bar{z})T(0,0)\right>, 
  \:\:
\bar{\partial}\left<T(z,\bar{z})\Theta(0,0)\right> = -\partial\left<\Theta(z,\bar{z})\Theta(0,0)\right>.\label{t-th}
\end{eqnarray}
We  note that we can generally write these correlators as
\begin{eqnarray}
\left<T(z,\bar{z})T(0,0)\right> = \frac{F(z\bar{z})}{z^4},\quad
  \left<T(z,\bar{z})\Theta(0,0)\right> =
  \frac{G(z\bar{z})}{z^3\bar{z}}, \quad
  \left<\Theta(z,\bz)\Theta(0,0)\right> = \frac{H(z\bar{z})}{z^2\bar{z}^2}.
\end{eqnarray}
Substituting these into (\ref{t-th}), we find
$t\frac{d}{dt}F(t)=3G(t)-t\frac{d}{dt}G(t)$ and 
$t\frac{d}{dt}G(t)= G(t)+2H(t)-t\frac{d}{dt}{H}(t)$ , 
where we define a Lorentz invariant scale parameter
$t=z\bar{z}$.
Combining these two equations, we
can prove a function defined as
\begin{eqnarray}
c_{L}(t)=2F(t)-4G(t)-6H(t)
\end{eqnarray}
is monotonically non-increasing along the RG flow, i.e., 
\begin{eqnarray}
 t\frac{d}{dt}c_L(t) = 
 -12H(t)\leq 0.
 \label{decreaseleft}
\end{eqnarray}
 Note that   
  this $c$-function has its extremum at the fixed points because 
   the trace of the stress tensor $\Theta=T_{z\bar{z}}$ 
vanishes when the theory has conformal invariance. 
The value of the $c$-function at the fixed points is equal to the 
central charge of the left-moving Virasoro algebra.

In the similar way, by using (\ref{T-conserv}) we
can show that another function
\begin{eqnarray}
c_{R}(t)=2\bar F(t)-4\bar G(t)-6H(t)
\end{eqnarray}
is 
also monotonically non-increasing, 
\begin{eqnarray}
 t\frac{d}{dt}c_R(t) =
  -12H(t) \leq 0.
  \label{decreaseright}
\end{eqnarray}
Here $\bar F(t)$ and $\bar G(t)$ are the right-moving counterparts of $F(t)$  
and $G(t)$, respectively.
From (\ref{decreaseleft}) and (\ref{decreaseright}), 
we can confirm  (\ref{eq.1}). 
In a parity symmetric theory, 
$c_L(t)$  is equal to $c_R(t)$
since in terms of the complex coordinate the parity symmetry means the
symmetry that exchanges $z$ and $\bar{z}$. In a parity violating case,
however, the  two c-functions,  $c_L(t)$ and $c_R(t)$,  differ from each 
other in general,  but (\ref{eq.1}) claims that the difference is a constant.  



We now briefly discuss the relation of the $c$-functions to the Weyl and
gravitational anomalies
\footnote{
From here, we move from Euclidean to Minkowski signature.
}. In order to analyse these, 
here we consider the two dimensional field theory coupled to some curved background.

It is well-known that, at the fixed point of the RG flow,
the Weyl anomaly is expressed as
\begin{eqnarray}
 \left<T^i_{\hspace{2mm}i}\right> &=& \frac{1}{24\pi}\frac{c_L+c_R}{2}R,\label{Weyl-anomaly}
\end{eqnarray}
in terms of the sum of the two central charges. 
It is related to the number of dynamical degrees of freedom of fluctuating 
fields. 
Away from the fixed point, the Weyl anomaly (\ref{Weyl-anomaly}) receives
corrections proportional to beta functions. But nevertheless, the 
coefficient of the scalar curvature must be still related to the 
effective degrees of freedom. Hence, it is natural to regard the coefficient  
as {\it the sum of the two $c$-functions}, $c_{L}(t)+c_{R}(t)$, along  the RG flow.

Now let us discuss the gravitational anomaly. At the fixed point, it is
expressed as a violation of momentum conservation
\begin{eqnarray}
 \nabla_{i}\left<T^{ij}\right> &=& -\frac{c_L-c_R}{96\pi}
 \epsilon^{jk}\partial_{k}R.\label{Grav-anomaly}
\end{eqnarray}
The gravitational
anomaly occurs due to the lack of a regularization which preserves
general covariance, and the general covariance of the quantum action is
broken by parity-violating one-loop diagrams \cite{AW}. This is
proportional to the difference between two central charges, and thus we obtained the above expression.
Away from the fixed point, in contrast to the Weyl anomaly, 
Eq.~(\ref{Grav-anomaly}) does not
receive any corrections. For the same reason as before, it is natural to
identify the coefficient of the
r.h.s.  of (\ref{Grav-anomaly}) with {\it the difference of the two $c$-functions}, $c_{L}(t)-c_{R}(t)$, along  the RG flow.

We finally note that, in the presence of the gravitational anomaly,
the vacuum expectation value of general  energy-momentum tensors 
is not covariant. 
The energy-momentum tensor 
in  Eqs. (\ref{Weyl-anomaly}) and
(\ref{Grav-anomaly}) is therefore understood as modified  
so as to be covariant 
 by adding  the so-called
Bardeen-Zumino polynomial \cite{BZ}(for useful review, see \cite{Bertlmann}).


\section{Holography with gravitational  Chern-Simons term}

In this section, we consider holographic RG flow of the three dimensional gravity which includes
the gravitational Chern-Simons term. 
We define two different $c$-functions, one for the left-mover and another for the
right-mover, the difference of which comes from the existence of the gravitational anomaly.

\subsection{ADM decomposition and canonical formalism}

We consider the three dimensional gravity with gravitational Chern-Simons term coupled to some scalar fields $\phi^I$.
The action is given by
\begin{eqnarray}
 \frac{1}{16\pi G_{\rm{N}}} \int  d^3x {\mathcal L} &=& 
 \frac{1}{16\pi G_{\rm{N}}} \int d^3x \sqrt{-\hat{g}}\bigg\{ \hat{R} 
 - \frac{1}{2} G(\phi)_{IJ} \partial _\mu \phi^I \partial ^\mu \phi^J -V(\phi)
 \nonumber \\[1ex]
&&
 + \frac{\beta}{2} \hat{\epsilon}^{\mu\nu\rho} 
 \Big( \hat{\Gamma}^\alpha{}_{\mu\beta} \partial_\nu \hat{\Gamma}^\beta{}_{\rho\alpha} +
 \frac{2}{3}\hat{\Gamma}^\alpha{}_{\mu\beta} \hat{\Gamma}^\beta{}_{\nu\gamma} 
 \hat{\Gamma}^\gamma{}_{\rho\alpha}\Big) \bigg\}, \label{eq:action}
\end{eqnarray}
where 
$G_{\rm{N}}$ is the three dimensional Newton constant 
and the coefficient $\beta$ is some constant. 
The hat is used for the quantities constructed out of 
the metric $\hat{g}_{\mu\nu}$, and
Greek indices run over the three dimensional coordinates $(r,x^0,x^1)$. 

Let us parametrize the metric as
\begin{eqnarray}
 ds^2 = \hat{g}_{\mu\nu}dx^\mu dx^\nu = N^2dr^2 + g_{ij}\left(dx^i +
							 N^idr\right)\left(dx^j
							 + N^jdr\right),
\end{eqnarray}
where $g_{ij}$ denotes the two dimensional boundary metric and Latin indices run over
the two dimensional coordinates $(x^0,x^1)$. 
Then we perform the Euclidian ADM-decomposition of the Lagrangian as follows:
\begin{alignat}{3}
  {\mathcal L} &= \sqrt{-g} N\left(R - V(\phi) - K_{ij}K^{ij} + K^2\right) \notag 
  \\
  &\quad\,
  - \frac{\sqrt{-g}}{2N} G_{IJ} \left(\dot{\phi}^I - N^i\partial_i\phi^I\right) 
  \left(\dot{\phi}^J - N^i\partial_i\phi^J\right)
  -\frac{\sqrt{-g}N}{2} G_{IJ} \partial_i\phi^I \partial^i\phi^J \notag
  \\
  &\quad\,
  +\beta\sqrt{-g}\epsilon^{nm}\dot{K}_{mk}K_n{}^{k}
  +\beta\sqrt{-g}N\left(2\epsilon^{mn}\nabla_{k}\nabla_{n}K_{m}{}^{k} -A^{mn}K_{mn}\right) \notag
  \\
  &\quad\,
  +\beta\sqrt{-g}N^{i}\left[2\epsilon^{mn}K_i{}^k\nabla_{n} K_{mk} +
  \epsilon^{mn}\nabla_{k}\left(K_{ni}K_{m}{}^k \right) 
  +\frac{1}{2}\epsilon_{ij}\partial^j R + \nabla_k A_{i}^{\h k}\right], \label{adm}
\end{alignat}
where total derivatives are omitted. Here the boundary metric
$g_{ij}$ is used in order to raise or lower Latin indices.
We use a dot as $\partial/\partial r$ and
$\epsilon^{ij}$ denotes the covariantly constant anti-symmetric
tensor. In terms of $N,N^i$ and $g_{ij}$, the extrinsic curvature can be
written as
\begin{alignat}{3}
 K_{ij} = \frac{1}{2N}\left(\dot{g}_{ij} -
			 \nabla_iN_j - \nabla_jN_i\right)
			 \label{extrinsic}, 
\end{alignat}
and $K$ is the trace part of $K_{ij}$.
$A_{ij}$ is explicitly written as
\begin{alignat}{3}
 A^{ij} &= \epsilon^{mp}g^{lo}T^{ijk}_{mno}\nabla_k\Gamma^n_{\h pl}, 
  \quad \quad T^{ijk}_{mno} :=
  \frac{1}{2}\left(\delta^k_{m}\delta^{(i}_o\delta^{j)}_n +
	      \delta^{k}_n\delta^{(i}_o\delta^{j)}_m - \delta^{k}_o\delta^{(i}_m\delta^{j)}_n\right), 
\end{alignat}
and does not behave as a tensor under diffeomorphisms on the boundary because it depends on the affine
connection in an explicit way.

In order to construct the Hamilton-Jacobi equation of the system, we
consider the canonical formalism with the gravitational 
Chern-Simons term
\cite{HHKT}. 
Since 
$K_{ij}$ has an  $r$-derivative term as in 
(\ref{extrinsic}), the action (\ref{adm}) contains third derivatives with
respect to $r$. 
It is known that the canonical formalism of such a system
is formulated 
by using (modified) Ostrogradsky method \cite{Ostrogradsky, BL, BLK} in which
a Lagrange multiplier is introduced and the extrinsic curvature
$K_{ij}$ is treated as an independent variable.
A close look at (\ref{adm}) shows that 
${\cal L}$ contains third derivatives of only the traceless part of
$K_{ij}$ and there is no   $\dot K$ term. We therefore
divide $K_{ij}$ into the trace part $K$ and traceless part $H_{ij} =
K_{ij}-g_{ij}K/2$.
The Lagrangian (\ref{adm}) can be rewritten as
\begin{alignat}{3}
  {\mathcal L} &= \sqrt{-g} N \left( R - V(\phi) - H_{ij}H^{ij} + \frac{1}{2}K^2 \right) \notag
  \\
  &\quad\,
  - \frac{\sqrt{-g}}{2N} G_{IJ} \left(\dot{\phi}^I - N^i\partial_i\phi^I\right) 
  \left(\dot{\phi}^J - N^i\partial_i\phi^J\right)
  -\frac{\sqrt{-g}N}{2} G_{IJ} \partial_i\phi^I \partial^i\phi^J \notag
  \\
  &\quad\, 
  + \beta\sqrt{-g}\dot{H}_{mk}\epsilon^{n(m}H^{k)}_{\hh n} 
  + \frac{1}{2}\beta\sqrt{-g}K\dot{g}_{mk}\epsilon^{n(m}H^{k)}_{\hspace{2mm}n}
  + \beta\sqrt{-g}N\left(2\epsilon^{mn}\nabla_{k}\nabla_{n}H_{m}^{\h k} - A^{mn}\textcolor{black}{K_{mn}}\right) \notag
  \\
  &\quad\,  
  + \beta\sqrt{-g}N^{i}\left\{2\epsilon^{mn}\textcolor{black}{K_i^{\h k}\nabla_{n} K_{mk}} +
  \epsilon^{mn}\nabla_{k}\left(\textcolor{black}{K_{ni}K_{m}^{\h k}}\right) +
  \frac{1}{2}\epsilon_{ij}\partial^j R + \nabla_k A_{i}^{\h k}\right\} \notag 
  \\ 
  &\quad\,
  + v^{ij}\left(\dot{g}_{ij}-2N\textcolor{black}{K_{ij}}- 2\nabla_{i}N_j\right), 
  \label{eq;calL}
\end{alignat}
Here $v^{ij}$ is a Lagrange multiplier. We treat $g_{ij}$, $H_{ij}$  and
$K$ as independent variables and $K_{ij}$ in 
(\ref{eq;calL}) is understood as $K_{ij}=H_{ij}+g_{ij}K/2$. 
Note that $v^{ij}$ is symmetric but not a tensor.

From this expression, we read off the momenta conjugate to
${g}_{ij}, {H}_{ij}$ and $\phi^I$ as
\begin{eqnarray}
 \pi^{ij} &:=& \frac{\delta {\mathcal L}}{\delta \dot{g}_{ij}} =
  v^{ij}+\frac{K}{2}\beta\sqrt{-g}\epsilon^{k(i}H^{j)}_{\hspace{2mm}k} ,
\\
\Pi^{ij} &:=& \frac{\delta {\mathcal L}}{\delta \dot{H}_{ij}} =
 \beta\sqrt{-g}\epsilon^{k(i}H^{j)}_{\hh k},
\label{eq5}\\
\pi_I &:=& \frac{\delta {\mathcal L}}{\delta \dot{\phi}^I} = 
- \frac{\sqrt{-g}}{N} G_{IJ }\left(\dot{\phi}^J-N_i \partial ^i\phi^J \right),
\end{eqnarray}
respectively. Now we find that $\Pi^{ij}$ is not independent of
$H_{ij}$ and the system is constrained.  Again, such a kind of the
constraint should be taken into account by introducing a Lagrange
multiplier $f_{ij}$, 
when we treat $H_{ij}$ and $\Pi^{ij}$ as independent variables. 
Hence, we add the following term 
\begin{eqnarray}
 f_{ij}\left(\beta\sqrt{-g}\epsilon^{k(i}H^{j)}_{\hh k}-\Pi^{ij}\right).
\end{eqnarray}
Then the total action can be written as
\begin{alignat}{3}
&S\left[g_{ij},H_{ij}, K, \phi^I,\pi^{ij},\Pi^{ij},\pi_I,N,N^i,f_{ij}; r_0\right]
\label{action} \\
&\hspace{-0mm}=
 \int d^2x\int^{r_0}\!\!dr \left[\pi^{ij}\dot{g}_{ij} 
+ \Pi^{ij}\dot{H}_{ij} +
  \pi_I\dot{\phi}^I\right.
\left. 
-\left(N{\mathcal H} +N_i{\mathcal P}^i\right) + f_{ij}\left(
		      \beta\sqrt{-g}\epsilon^{k(i}H^{j)}_{\hh
		      k}-\Pi^{ij}\right)\right],
		      \nonumber 
\end{alignat}
where $r$-integration has been cut off at $r=r_{0}$.
Here we introduced 
{\em Hamiltonian} and {\em momentum} as follows:
\begin{alignat}{3}
 \frac{1}{\sqrt{-g}} {\mathcal H} &:= -R + V(\phi) + H^{kl}H_{kl} - \frac{1}{2}K^2 
  - \frac{1}{2(-g)}G^{IJ}\pi_I \pi_J + \frac{1}{2} G_{IJ} \partial_i\phi^I \partial^i\phi^J \nonumber 
  \\
  &\quad\;
  - 2\beta\epsilon^{mn}\nabla_k\nabla_n H_m^{\h k} + \left(\frac{2}{\sqrt{-g}}\pi^{kl} 
  + \beta A^{kl}\right)\left(H_{kl} + \frac{1}{2}g_{kl}K\right),
  \\[2ex]
\frac{1}{\sqrt{-g}}  {\mathcal P^i} &:= -2\beta\epsilon^{mn}\textcolor{black}{K^{i k}\nabla_n K_{mk}} 
  -\beta\epsilon^{mn}\nabla_k\left(\textcolor{black}{K_{n}^{\h i}K_m^{\h k}}\right) 
  +\beta\nabla_j\left(K\epsilon^{k(i}H^{j)}_{\hspace{2mm}k}\right) \nonumber 
  \\
  &\quad\;
  -\frac{1}{2}\beta\epsilon^{ij}\partial_j R
  -\nabla_j\left(\frac{2}{\sqrt{-g}}\pi^{i j} + \beta A^{i j}\right) 
  + \frac{1}{\sqrt{-g}}\pi_I \partial ^i\phi^I.
\end{alignat}
We find $N, N_i, f_{ij}$ and $K$ are 
Lagrange multipliers in (\ref{action}). Path integrations over 
them lead to the following constraints:
\begin{eqnarray}
 &&{\mathcal H} = 0,  \quad\quad {\mathcal P}^i = 0, \quad \quad 
 \label{constraints0}
 \\
 [2mm]
 && \Pi^{ij} =
  \beta\sqrt{-g}\epsilon^{k(i}H^{j)}_{\hh k},
  \label{constraints1} 
  \\[2mm]
 && 
K=\left ( \frac{1}{\sqrt{-g}}\pi ^{ij}+\frac{\beta }{2}A^{ij}
\right )g_{ij}
\label{constraints2}
\end{eqnarray}

\subsection{Hamilton-Jacobi equation}

Now we consider the Hamilton-Jacobi equation in the bulk. We have six
dynamical fields,  $g_{ij}, H_{ij}, \phi^I, \pi^{ij}, \Pi^{ij}$ and  $\pi_I$, 
together with  
four auxiliary fields,  $N, N_i, K$ and  $f_{ij}$. 
In order to obtain a classical
solution, 
$\bar g_{ij}(r, x)$, e.t.c., 
we have to give the boundary conditions for dynamical fields
which are consistent with the constraints 
(\ref{constraints0}), 
(\ref{constraints1}) and (\ref{constraints2}). 
Since we are interested in the variations of fields along the
radial direction, we give the following boundary conditions:
\begin{eqnarray}
\bar g_{ij}(r_0,x) = g_{ij}(x),\quad 
\bar H_{ij}(r_0,x) = H_{ij}(x), \quad 
\bar \phi^I(r_0,x)=\phi^I(x),
\end{eqnarray}
where $r=r_0$ denotes the two-dimensional boundary surface
\footnote{To deal with second-order differential equations, 
we of course need to give one more boundary condition
for each field, which is assumed to be fixed so that the classical
solution is regular inside the bulk \cite{Maldacena, GKP, Witten}(see
also Ref. \cite{Graham-Lee}). 
}.

We denote as $\bar{S}$ the action in which the classical solution is
substituted. Clearly this is a functional of boundary conditions
$g_{ij}(x), K_{ij}(x)$ and $\phi^ I(x)$, and should be written as
$\bar{S}[g_{ij}, H_{ij}, \phi^ I; r_0]$. 
Since the classical solution 
satisfies the constraints
(\ref{constraints1}), the
classical action $\bar{S}$ can be written explicitly as
\begin{eqnarray}
 \bar{S}\left[g_{ij},H_{ij},\phi^I  ;r_0\right] = \int d^2x
  \int^{r_0}\!\!dr \left(\bar {\pi}^{ij}\dot{\bar {g}}_{ij}+
		    \bar {\Pi}^{ij}\dot{\bar{H}}_{ij} + 
		    \bar {\pi}_I \dot{\bar{\phi}}^I \right). \label{eq:S}
\end{eqnarray}
Then the variation of the action can
be written as
\begin{eqnarray}
 \delta\bar{S}\left[g_{ij},H_{ij},\phi^I ; r_0\right]
&=&  \frac{\partial S}{\partial r_0} \delta r_0 +  \int d^2x \left[\pi^{ij}\delta g_{ij} +
	       \Pi^{ij}\delta H_{ij} +
	       \pi_I  \delta\phi^I  \right],
\end{eqnarray}
where only surface terms contribute to $\delta \bar{S}$ due 
to the classical equations of motion. 
Here $\pi _{ij}$ denotes $\bar \pi _{ij}(r_{0}, x)$ and similarly 
for $\Pi _{ij}$ and $\pi_I $. 
We find from this
expression that
\begin{eqnarray}
 \frac{\delta \bar{S}}{\delta
 g_{ij}} = \pi^{ij},\quad \frac{\delta \bar{S}}{\delta H_{ij}} =
 \Pi^{ij},\quad \frac{\delta \bar{S}}{\delta \phi^I} = \pi_I, \quad
 \frac{\partial \bar{S}}{\partial r_0} = 0. \label{H-J}
\end{eqnarray}
The last equation is obtained by the total differentiation of Eq.~(\ref{eq:S})
with respect to $r_0$, and shows that  $\bar{S}$ is
independent of the position of the boundary surface. The
classical action is therefore determined by the boundary conditions
$g_{ij}, H_{ij}$ and $\phi^I$, which are constrained by the
Hamiltonian one, $\mathcal{H}=0$, the momentum one, $\mathcal{P}^i = 0$, and the others,  Eqs.~(\ref{constraints1}) and (\ref{constraints2}).

\subsection{Holographic RG equation}

Let us now discuss the physical meaning of the Hamiltonian constraint
${\mathcal H}=0$. By using the relations of (\ref{H-J}), we can rewrite it as
\begin{eqnarray}
& & \hskip-1.5cm
 \frac{1}{2}G^{IJ}\left(\frac{1}{\sqrt{-g}}\frac{\delta \bar{S}}{\delta \phi^I}\right)\left(\frac{1}{\sqrt{-g}}\frac{\delta \bar{S}}{\delta \phi^J}\right)
 -
		   \left(\frac{2}{\sqrt{-g}}\frac{\delta \bar{S}}{\delta g_{kl}} + 
		   \beta
		    A^{kl}\right)\left(H_{kl} + \frac{1}{2}g_{kl}K\right) 
		    \nonumber \\
		    &=&  -R+V(\phi)
		   +\frac{1}{2}g^{ij}G_{IJ}\partial _i\phi^I\partial _j\phi^J
 +H^{kl}H_{kl}-\frac{1}{2}K^2-2\beta\epsilon^{mn}\nabla_k\nabla_n
  H_m^{\h k}.
\end{eqnarray}
This is the extension of the flow equation of de Boer, Verlinde and Verlinde \cite{BVV} 
by the inclusion of the gravitational Chern-Simons term\footnote{Inclusion of higher derivative terms
with even parity is investigated in refs.~\cite{Fukuma2}.}.
From this flow equation, we can determine the classical action $\bar{S}$ as a
functional of boundary variables, $g_{ij}, H_{ij}$ and $\phi^I$. To see
this, let us assign the weight $w$ as
\begin{eqnarray}
 w = 0&:&\,\, g_{ij},\,\, H_{ij},\,\, \phi^I,\,\, \Gamma[g,H,\phi^I],
 \nonumber \\[0.2cm]
 w = 1&:&\,\, \partial_i,\\
 w = 2&:&\,\, R,\,\, \partial_i\phi^I\partial_j\phi^J, \:\:
 \frac{\delta \Gamma }{\delta g_{ij}},\:\:\frac{\delta \Gamma }{\delta H_{ij}},
 \:\:\frac{\delta \Gamma }{\delta \phi ^{I}}, \:\:\: A^{ij}, 
 \nonumber 
\end{eqnarray}
and expand $\bar{S}$ as
\begin{eqnarray}
 \bar{S} &=& 16\pi G_{\rm N} \Gamma + \sum_{w=0}^\infty S_{\rm loc}^{(w)},
\end{eqnarray}
where $S_{\rm loc}^{(w)}$ contains local terms with weight $w$ and
only $\Gamma$ has non-local terms. In the context of the holographic RG, the non-local part $\Gamma[g,H,\phi]$ can be regarded as
the generating functional with respect to the source fields.

Hereafter through the end of this subsection, 
we assume $H_{ij}=0$ 
for simplicity. 
Let us recall in this connection that 
 ordinary solutions such as global AdS$_3$ 
and BTZ black hole  can
satisfy this assumption by performing suitable coordinate 
transformations, although it is an interesting task  to relax this assumption.
Under the assumption, we find that the Hamiltonian constraint reduces to
\begin{alignat}{3}
  &-\frac{1}{2g}G^{IJ}\frac{\delta \bar{S}}{\delta \phi^I}\frac{\delta
   \bar{S}}{\delta \phi^J}
 -\frac{1}{2}\left \{ 
 \Big( \frac{1}{\sqrt{-g}}\frac{\delta \bar{S}}{\delta g_{kl}} 
 + \frac{\beta}{2}A^{kl} \Big) g_{kl}\right \}^{2} 
 =  -R+V(\phi)
		   +\frac{1}{2}g^{ij}G_{IJ}\partial_i\phi^I\partial_j\phi^J .\label{flow}
\end{alignat}
Here use has been made of (\ref{constraints2}). 
We now solve the flow equation (\ref{flow}) order by order with respect to
weight $w$. For $w=0$, defining ``superpotential'' $W(\phi )$ by 
$S_{\rm loc}^{(0)} = \int d^2x\sqrt{-g}W(\phi )$, we find
\begin{eqnarray}
 V &=& \frac{1}{2}G^{IJ}\frac{\partial W}{\partial \phi^I}\frac{\partial
  W}{\partial \phi^J} - \frac{1}{2}W^2. \label{eq:superpot}
\end{eqnarray}
This is an equation to determine $W$, given the potential  $V$.

Next, we consider the flow equation for $w=2$.  
Since the scalar fields are regarded as coupling constants of 
the dual field theory, we 
set  $\phi$ to be  constant on  the two-dimensional surface. 
Then the $w=2$ flow equation turns out to be 
\begin{eqnarray}
  g_{ij}\left(\frac{2}{\sqrt{-g}}\frac{\delta \Gamma}{\delta
	 g_{ij}}+\frac{\beta}{16\pi G_{\rm N}} A^{ij} \right) - G^{IJ}\frac{2}{W}\frac{\partial W}{\partial
 \phi^J}\left(\frac{1}{\sqrt{-g}}\frac{\delta \Gamma}{\delta \phi^I}\right)
 &=& \frac{1}{16\pi G_{\rm N}}\frac{2}{W}R 
  .
  \label{weight-two}
\end{eqnarray}
If 
the two-dimensional metric is flat, 
 this equation leads to the holographic RG equation  
of the dual field theory, 
and  the holographic beta function is defined as 
\begin{eqnarray}
\beta^I(\phi) =G^{IJ}\frac{2}{W}\frac{\partial W}{\partial\phi^J}. 
\label{eq:beta}
\end{eqnarray}
Putting (\ref{eq:beta}) back into (\ref{weight-two}), we can 
regard (\ref{weight-two}) as the Weyl anomaly
equation of the dual field theory on the curved backgrounds:
\begin{eqnarray}
 \left<T^i_{\hspace{2mm}i}\right> &=& \frac{1}{24\pi } 
 \frac{3}{G_{\rm N}W}R +
  \beta^I(\phi)\frac{1}{\sqrt{-g}}\frac{\delta \Gamma}{\delta \phi^I}  .
\label{eq:traceanomaly}
\end{eqnarray}
Here the covariant energy-momentum tensor $T^{ij}$ is understood as modified 
by adding the Bardeen-Zumino term, i.e., 
$\beta A^{ij}/16\pi G_{\rm N}$ to the ordinary one 
$(2/\sqrt{-g})\delta \Gamma /\delta g_{ij}$. 
At the fixed points where $\beta^I (\phi )=0$ in the second term 
of the r.h.s of (\ref{eq:traceanomaly}), we can read off the 
sum of the central charges of the left- and right-movers from the 
coefficients of the scalar curvature. 
Away from the fixed points, it is legitimate to  define the sum of
$c$-functions for left- and right-movers in the dual field theory as
follows:
\begin{eqnarray}
 c_L(\phi) + c_R(\phi) &=& 
 \frac{6}{G_{\rm N}W(\phi)}.
 \label{L+R}
\end{eqnarray}

\subsection{Gravitational Anomaly}

We then discuss the physical meaning of the momentum constraint
${\mathcal P}^i = 0$. By
using the relations (\ref{H-J}), the momentum constraint can be written as
\begin{eqnarray}
 \frac{1}{2}\beta\sqrt{-g}\xi_i\epsilon^{ij}\partial_jR &=&
  \xi_i\left\{-2\beta\sqrt{-g}\epsilon^{mn}K^{il}\nabla_nK_{ml}
  -\beta\sqrt{-g}\epsilon^{mn}\nabla_k\left(K_{n}^{\hspace{2mm}i}K_m^{\hspace{2mm}k}\right)\right.
  \nonumber \\
&& 
\hskip-1cm 
\left.+\beta\sqrt{-g}\nabla_j\left(K\epsilon^{k(i}H^{j)}_{\hspace{2mm}k}\right)- \nabla_j\left(2\frac{\delta \bar{S}}{\delta g_{ij}}+\beta\sqrt{-g}
	      A^{ij}\right) +
\partial^i\phi^I\frac{\delta\bar{S}}{\delta\phi^I}\right\}.
\end{eqnarray}
Up to total derivatives, this is equivalent to
\begin{eqnarray}
 -\frac{\beta}{4}\sqrt{-g}\xi_i\epsilon^{ij}\partial_j R 
 &=& \delta H_{ij}\frac{\delta\bar{S}}{\delta H_{ij}} + \delta
  g_{ij}\left(\frac{\delta\bar{S}}{\delta g_{ij}} + \frac{1}{2}\beta\sqrt{-g} A^{ij}\right) + \delta\phi^I\frac{\delta
  \bar{S}}{\delta \phi^I},\label{grav-anomaly}
\end{eqnarray}
where $\delta$ means the infinitesimal field variation with respect to the
general coordinate transformation $x^i\to x^i+\xi^i$. In order to derive
this expression, we use
\begin{eqnarray}
&& \xi_i\left\{-2\beta\sqrt{-g}\epsilon^{mn}K^{il}\nabla_nK_{ml} -
       \beta\sqrt{-g}\epsilon^{mn}\nabla_k\left(K_{n}^{\hspace{2mm}i}K_m^{\hspace{2mm}k}\right)
	+\beta\sqrt{-g}\nabla_j\left(K\epsilon^{k(i}H^{j)}_{\hspace{2mm}k}\right)\right\}
\nonumber \\
&=& \sqrt{-g}\xi_i\Big\{-\beta\epsilon^{mn}K_{n}^{\hspace{2mm}l}\nabla^iK_{ml}
	  +\beta\epsilon^{mn}\nabla_m\left(K^i_{\hspace{2mm}l}K_{n}^{\hspace{2mm}l}\right)
	 \nonumber \\
&&\hspace{5cm}+\beta\epsilon^{mn}\nabla_l\left(K^i_{\hspace{2mm}m}K_n^{\hspace{2mm}l}\right)+\beta\nabla_j\left(K\epsilon^{k(i}H^{j)}_{\hspace{2mm}k}\right)\Big\}
	\nonumber \\
&=& -\xi_i\Pi^{mn}\nabla^iK_{mn} +
 2\xi_i\nabla_m\left(K^i_{\hspace{2mm}n}\Pi^{mn}\right) + \xi_i\nabla_j
 \left(K\Pi^{ij}\right)
 \nonumber \\
&\simeq& -\left(\xi^i\nabla_iK_{mn} + 2K_{in}\nabla_m\xi^i 
-K\nabla_m\xi_n\right)\Pi^{mn},
\end{eqnarray}
and the 
constraint of (\ref{constraints1}).

Eq. 
(\ref{grav-anomaly}) means that the classical bulk action $\bar{S}$ is
not invariant under two-dimensional diffeomorphisms. More explicitly, by
extracting $w=3$ terms of both sides and set $\phi$ to be constant on the
two-dimensional surface, we obtain 
\begin{eqnarray}
 \nabla_i \left<T^{ij}\right>  = -\frac{3\beta}{G_{\rm N}}\frac{1}{96\pi}\epsilon^{ij}\partial_jR,
\end{eqnarray}
which implies that, in
the two-dimensional dual field theory, there is a gravitational anomaly proportional to the Chern-Simons
coupling $\beta$. Hence, we can {\em define  the difference between two
$c$-functions for left-and right-mover} as
\begin{eqnarray}
 c_L(\phi) - c_R(\phi) &=& \frac{3\beta}{G_{\rm N}}\label{L-R}.
\end{eqnarray}
Combining (\ref{L+R}) and (\ref{L-R}), we finally obtain the holographic
expression of the left-right asymmetric $c$-functions:
\begin{eqnarray}
 c_L(\phi) = \frac{3}{G_{\rm
  N}}\left(\frac{1}{W(\phi)}+\frac{\beta}{2}\right),\quad c_R(\phi) =
  \frac{3}{G_{\rm N}}\left(\frac{1}{W(\phi)}-\frac{\beta}{2}\right).
  \label{eq:finalcfunctions}
\end{eqnarray}
As we confirm in Appendix 
\ref{sec:example} by examples, these $c$-functions are 
both monotonically non-increasing toward the infrared direction.
If the scalar potential $V$ is just a negative constant, i.e.,
\begin{eqnarray}
V=-\frac{2}{\ell^{2}}
\end{eqnarray} 
then we find $W(\phi)=2/\ell$ and 
(\ref{eq:finalcfunctions}) turns out to be the central charges 
\begin{eqnarray}
c_L = \frac{3}{G_{\rm
  N}}\left(\frac{\ell}{2}+\frac{\beta}{2}\right),\quad c_R =
  \frac{3}{G_{\rm N}}\left(\frac{\ell}{2}-\frac{\beta}{2}\right)
\end{eqnarray}
which we obtained previously by the Brown-Henneaux's method \cite{HHKT}.

\section{Conclusion and discussion}

In this paper, we investigated the three dimensional gravity 
with gravitational Chern-Simons term 
coupled to some scalar fields.
This theory admits a solution which interpolates two AdS$_3$ vacua due 
to the nontrivial
profile of the scalar fields in the radial direction. 
The solution can be interpreted as the holographic RG 
flow of the two dimensional boundary quantum field theory.
We constructed $c$-functions for left and right movers, which are consistent 
with Zamolodchikov's $c$-theorem.
The presence of the gravitational Chern-Simons term is crucial to obtain the
left-right asymmetric ones.

First we decomposed the radial direction of the three dimensional 
theory in an ADM manner and applied the canonical method.
Since the action contains the third derivative term with respect to the radial coordinate,
we introduced the Lagrange multiplier and identified the extrinsic curvature as an 
independent variable.
From this prescription we obtained 
the Hamiltonian and momentum constraints.

By inserting the classical solution with arbitrary boundary variables to the action, 
we obtained the Hamilton-Jacobi equation from the Hamiltonian constraint.
This equation is solved order by order with respect to the weight $w$,
and the holographic RG flow equation is obtained for the case of $w=2$.
From this, the beta functions for the scalar fields are derived.
Furthermore, by making a comparison  with the Weyl anomaly 
on the boundary field theory, 
the sum of the left and right $c$-functions, $c_L(\phi) + c_R(\phi)$, 
is  obtained.

On the other hand, the momentum constraint is compared with the gravitational 
anomaly of the two dimensional quantum field theory.
From this, it is possible to read the difference of the two $c$-functions, $c_L(\phi) - c_R(\phi)$, which is constant along the RG flow. 
That the difference is constant is in perfect agreement with the 
$c$-theorem for parity violating two-dimensional quantum field theories.

In conclusion, by analyzing the three dimensional gravity with gravitational Chern-Simons term
coupled to some scalar fields, we obtained the left-right asymmetric $c$-functions.
These are monotonically non-increasing along the RG flow toward the IR region and
precisely agree with the central charges at the fixed points.
These results confirm the gauge/gravity correspondence in the presence of
the gravitational Chern-Simons term.

In Ref. \cite{chiralgravity}, it was shown  
in TMG that the AdS$_3$ vacuum is unstable against the perturbation 
of the gravitational fields except for the chiral case. 
As a future work, it is important to study the instability 
of our solution. 
It has been known that there exist stable solutions in TMG called 
warped AdS$_{3}$ \cite{alpss, anninos}. 
It is also an interesting task to study RG flow connecting  AdS$_3$ and 
warped AdS$_3$ or two warped AdS$_{3}$ vacua. We hope we could come to these 
problems in our future work.

\section*{Acknowledgements}

We would like to thank Kiyoshi Higashijima for useful comments.
KH and TN are supported in part by JSPS Research Fellowship for Young Scientists. 
The work of YH is partially supported by the Ministry of Education, Science, 
Sports and Culture, Grant-in-Aid for Young Scientists (B), 19740141, 2007.

\appendix
\section{Explicit form of the solution}\label{sec:example}

In this appendix, we discuss the holographic RG flow by solving the equations of motion for the action (\ref{eq:action}).
The equations of motion are written as
\begin{alignat}{3}
  0 &= \hat{R}_{\mu\nu} - \frac{1}{2} G_{IJ}(\phi) \partial_\mu \phi^I \partial_\nu \phi^J 
  - \hat{g}_{\mu\nu} V(\phi)
  + \beta \hat{\epsilon}_{\rho\sigma(\mu} \hat{\nabla}^\rho \hat{R}^\sigma_{\nu)}, \notag
  \\
  0 &= \hat{\nabla}_\mu ( G_{IJ} \hat{\partial}^\mu \phi^J) 
  - \frac{1}{2} \frac{\partial G_{JK}}{\partial \phi^I}  \partial_\mu \phi^J 
  \hat \partial^\mu \phi^K
  - \frac{\partial V}{\partial \phi^I}. \label{eq:eom}
\end{alignat}
By substituting the ansatz
\begin{alignat}{3}
  ds^2 = dr^2 + e^{-2 f(r)} \eta_{ij}dx^i dx^j, \quad \phi^I = \phi^I(r),
\end{alignat}
into (\ref{eq:eom}), we obtain a set of differential equations of the form,
\begin{alignat}{3}
  0 &= - \ddot{f} + \frac{1}{2} G_{IJ}(\phi) \dot{\phi}^I \dot{\phi}^J, \label{eq:eom1}
  \\
  0 &= - \ddot{f} + 2 \dot{f}^2 + V(\phi) , \label{eq:eom2}
  \\
  0 &= - \frac{\partial V}{\partial \phi^I} - 2 G_{IJ}(\phi) \dot{f} \dot{\phi}^J 
  + \frac{\partial G_{IJ}}{\partial \phi^K} \dot{\phi}^J \dot{\phi}^K 
  - \frac{1}{2} \frac{\partial G_{JK}}{\partial \phi^I} \dot{\phi}^J \dot{\phi}^K 
  + G_{IJ}(\phi) \ddot{\phi}^J. \label{eq:eom3}
\end{alignat}
Note that these equations do not depend on the coefficient $\beta$, and are the same as those  
for the Einstein gravity with the scalar field.
The third equation (\ref{eq:eom3}) can be derived by combining 
(\ref{eq:eom1}) and (\ref{eq:eom2}),
and hence neglected below.
By employing the superpotential and the $\beta$-function defined in 
Eqs.~(\ref{eq:superpot}) and (\ref{eq:beta}), 
we can rewrite Eqs.~(\ref{eq:eom1}) and (\ref{eq:eom2}) as follows:
\begin{alignat}{3}
  \dot{f} &= \frac{1}{2} W(\phi), \label{eq:ell}
  \\
  \dot{\phi}^I &= G^{IJ}(\phi) \frac{\partial W(\phi)}{\partial \phi^J}. \label{eq:phi}
\end{alignat}

In order to solve these equations, we consider the region $r_{\text{IR}} \leq r \leq r_{\text{UV}}$ and the solution where
$\dot{\phi}^I= 0$ only at $r=r_{\text{IR}}$ and $r_{\text{UV}}$. 
As discussed in Eq.~(\ref{L+R}), $W(\phi)$ is related to 
the $c$-function and hence should be positive.
Furthermore, 
if we regard $e^{f(r)}$  as the scale of boundary field theory, $t$, and use  
Eqs.~(\ref{eq:ell}) and (\ref{eq:phi}), we can identify the beta function $t\frac{d\phi^I}{dt}=G^{IJ}\frac{2}{W}\frac{\partial W}{\partial \phi^J}=\beta^I(\phi)$.
The fact that $\dot{\phi}^I=0$ is equivalent to  $\beta^I(\phi)=0$ implies that
$r_{\text{IR}}$ and $r_{\text{UV}}$ correspond to IR and UV fixed points in the sense of boundary  theory, respectively.
From these relations we obtain
\begin{alignat}{3}
  t\frac{d }{dt}c_L(\phi)= t\frac{d }{dt}c_R(\phi)=-\frac{3}{2G_\text{N}W(\phi)}\beta^I(\phi) G_{IJ}(\phi) \beta^J(\phi)\leq 0.
\end{alignat}
Therefore each $c$-function defined in Eq.~(\ref{eq:finalcfunctions}) is monotonically 
non-increasing along the RG flow.

Since Eqs.~(\ref{eq:ell}) and (\ref{eq:phi}) are just the first order differential equations, 
we can always find a solution for any superpotential $W(\phi)$ which satisfies the above conditions.
As an example, we consider only one scalar field $\phi \equiv \phi^1$ with the metric $G_{11}(\phi)=1$,
and choose the superpotential like
\begin{equation}
  W(\phi)=\frac{1}{\ell}\left(\sin\phi +\alpha \right), \label{super}
\end{equation}
where $\ell$ is some positive constant and $\alpha>1$.
This is monotonically non-decreasing in the region $-\pi/2 \leq \phi \leq \pi/2$.
Now the solution is given by
\begin{alignat}{3}
  f &= \frac{1}{2}\log(\cos\phi)-\alpha  \tanh ^{-1}\left( \tan \frac{\phi}{2}\right)+b, \notag
  \\
  \phi &= 2\arctan\left(  \tanh \left(\frac{r-a}{2\ell}\right)\right),
\end{alignat}
in the region $-\infty < r < \infty$.
Namely, the scalar field is represented by a kink solution.
The beta function is evaluated as
\begin{alignat}{3}
  \beta(\phi) = \frac{2\cos\phi}{\sin\phi+\alpha}.
\end{alignat}


\begin{thebibliography}{19}

\bibitem{DJ}
S. Deser and R. Jackiw,
``{\em  Three-Dimensional Cosmological Gravity: Dynamics of Constant Curvature,}''
 Annals Phys. {\bf 153} 405, (1984).
%
\bibitem{btz}
M. Banados, C. Teitelboim and J. Zanelli,
``{\em  The Black hole in three-dimensional space-time,}''
Phys. Rev. Lett. {\bf 69} 1849, (1992),  hep-th/9204099.
%
\bibitem{brown}
J. D. Brown and M. Henneaux,
``{\em  Central Charges in the Canonical Realization of Asymptotic Symmetries: An Example from Three-Dimensional Gravity,}''
Commun. Math. Phys. {\bf 104} 207, (1986).
%
\bibitem{strominger98}
A. Strominger, 
``{\it Black Hole Entropy from Near-Horizon Microstates,}" 
JHEP {\bf 9802} 009,  (1998), hep-th/971225.
%
\bibitem{HHKT} K. Hotta, Y. Hyakutake, T. Kubota, and H. Tanida,
	``{\em Brown-Henneaux's Canonical Approach to Topologically
	Massive Gravity},'' JHEP {\bf 0807}, 066 (2008) , arXiv:0805.2005 [hep-th].

\bibitem{DJT}
S. Deser, R. Jackiw and S. Templeton,
``{\em   Topologically Massive Gauge Theories,}''
Annals Phys. {\bf  140} 372, (1982), Erratum-ibid. {\bf  185} 406, (1988), Annals Phys. {\bf  281} 409, (2000).
``{\em  Three-Dimensional Massive Gauge Theories,}''
Phys. Rev. Lett. {\bf  48} 975, (1982).
\bibitem{chiralgravity}
W. Li, W. Song and A. Strominger,
``{\em Chiral Gravity in Three Dimensions,}''
JHEP {\bf 0804} 082 (2008), arXiv:0801.4566 [hep-th].

\bibitem{}
S. Carlip, S. Deser, A. Waldron and D.K. Wise,
``{\em  Cosmological Topologically Massive Gravitons and Photons,}'' Class. Quant. Grav. {\bf  26} 075008 (2009), arXiv:0803.3998 [hep-th].
%
\bibitem{}
D. Grumiller and N. Johansson, ``{\em  Instability in cosmological topologically massive gravity at the chiral point,}'' JHEP {\bf 0807} 134 (2008), arXiv:0805.2610 [hep-th].
%
\bibitem{}
G. Giribet, M. Kleban and M. Porrati, ``{\em  Topologically Massive Gravity at the Chiral Point is Not Chiral,}'' JHEP {\bf 0810} 045 (2008), arXiv:0807.4703 [hep-th]G
Y.-S. Myung, ``{\it Logarithmic conformal field theory approach to topologically massive gravity}h Physics Letters {\bf B670} 220 (2008), arXiv:0808.1942[hep-th].
%
\bibitem{mss}
A. Maloney, W. Song and A. Strominger,
``{\em  Chiral Gravity, Log Gravity and Extremal CFT,}'' arXiv:0903.4573 [hep-th].


\bibitem{alpss}
D. Anninos, W. Li, M. Padi, W. Song and A. Strominger,
``{\em  Warped AdS(3) Black Holes,}''
JHEP {\bf 0903} 130 (2009), arXiv:0807.3040 [hep-th].

\bibitem{anninos}
D. Anninos, M. Esole and M. Guica,
``{\em  Stability of warped AdS3 vacua of topologically massive gravity,}''
arXiv:0905.2612 [hep-th].
\bibitem{krauslarsen}
P. Kraus and F. Larsen,
``{\em  Microscopic black hole entropy in theories with higher derivatives,}''
JHEP {\bf 0509} 034, (2005), hep-th/0506176;
``{\em  Holographic gravitational anomalies,}''
JHEP {\bf 0601} 022, (2006), hep-th/0508218.





\bibitem{}
S. N. Solodukhin,
``{\em  Holography with gravitational Chern-Simons,}''
Phys. Rev. {\bf D74} 024015, (2006), hep-th/0509148.

\bibitem{}
Y. Tachikawa,
``{\em  Black hole entropy in the presence of Chern-Simons terms,}''
Class. Quant. Grav. {\bf 24} 737, (2007), hep-th/0611141.

\bibitem{park}
M.-I. Park,
``{\em  BTZ black hole with gravitational Chern-Simons: Thermodynamics and statistical entropy,}''
Phys. Rev. {\bf  D77} 026011, (2008), hep-th/0608165;
``{\em  BTZ Black Hole with Higher Derivatives, the Second Law of Thermodynamics, and Statistical Entropy: A New Proposal,}''
Phys. Rev.{\bf  D77} 126012, (2008), hep-th/0609027;
``{\em   Holography in Three-dimensional Kerr-de Sitter Space with a Gravitational Chern-Simons Term,}''
Class. Quant. Grav. {\bf  25} 135003, (2008), arXiv:0705.4381 [hep-th].

\bibitem{guptasen}
R. K. Gupta and A. Sen,
``{\em  Consistent Truncation to Three Dimensional (Super-) gravity,}''
JHEP {\bf  0803} 015, (2008), arXiv:0710.4177 [hep-th].

\bibitem{HHKNT}
K. Hotta, Y. Hyakutake, T. Kubota, T. Nishinaka and H. Tanida, 
``{\em  The CFT-interpolating Black Hole in Three Dimensions,}''
JHEP {\bf 0901} 010, (2009), arXiv:0811.0910 [hep-th].
\bibitem{Maldacena} J. Maldacena, ``{\em The large N limit of
	superconformal field theories and supergravity,}''
	Adv. Theor. Math. Phys. {\bf 2} (1998), 231, hep-th/9711200.
%
\bibitem{GKP} S. S. Gubser, I. R. Klebanov, and A. M. Polyakov, ``{\em
	Gauge Theory Correlators from Non-Critical String Theory,}''
	Phys. Lett. B{\bf 428} (1998), 105, hep-th/9802109.
%
\bibitem{Witten} E. Witten, ``{\em Anti De Sitter Space And
	Holography,}'' Adv. Theor. Math. Phys. {\bf 2} (1998), 253,
	hep-th/9802150.


\bibitem{BB}
G. Barnich and F. Brandt, ``Covariant theory of asymptotic symmetries, conservation laws and central charges,'' 
Nucl. Phys. B {\bf 633} (2002) 3, hep-th/0111246.

\bibitem{BC}
G. Barnich and G. Compere, ``Surface charge algebra in gauge theories and thermodynamic integrability,''
J. Math. Phys. {\bf 49} (2008) 042901, arXiv:0708.2378 [gr-qc].

\bibitem{ghss}
M. Guica, T. Hartman, W. Song and A. Strominger, ``The Kerr/CFT Correspondence,'' arXiv:0809.4266 [hep-th].

\bibitem{BVV} J. de Boer, E. Verlinde, and H. Verlinde, ``{\em On the
	Holographic Renormalization Group,}'' JHEP {\bf 0008}
	(2000),003,  hep-th/9912012.
\bibitem{Fukuma2} M. Fukuma, S. Matsuura, and T. Sakai, ``{\em A note on the
	Weyl anomaly in the holographic renormalization group},''
	Prog. Theor. Phys. {\bf 104} (2000), 1089, hep-th/0007062;
``{\em	Holographic Renormalization Group},'' Prog. Theor. Phys. {\bf
	109} (2003), 489, hep-th/0212314.
\bibitem{hotta}
K. Hotta,
``{\em  Holographic RG flow dual to attractor flow in extremal black holes,}''
Phys. Rev. {\bf D 79} 104018, (2009), arXiv:0902.3529 [hep-th].
\bibitem{fgpw}
D.Z. Freedman, S.S. Gubser, K. Pilch and N.P. Warner, 
``{\em   Renormalization group flows from holography supersymmetry and a c theorem,}''
Adv. Theor. Math. Phys. {\bf3} 363, (1999), hep-th/9904017.

\bibitem{st}
K. Skenderis and P.K. Townsend,
``{\em  Gravitational stability and renormalization group flow,}''
Phys. Lett. {\bf B468} 46, (1999), hep-th/9909070.

\bibitem{}
E.P. Verlinde and H.L. Verlinde, 
``{\em RG flow, gravity and the cosmological constant ,}''
JHEP {\bf 0005} 034, (2000), hep-th/9912018.

\bibitem{noo1}
S. Nojiri, S.D. Odintsov and S. Ogushi, 
``{\em   Conformal anomaly from d-5 gauged supergravity and c function away from conformity,}''
Grav. Cosmol. {\bf 6} 271, (2000), hep-th/9912191; 
``{\em   Holographic conformal anomaly with bulk scalars potential from d-3 and d-5 gauged supergravity,}''
Prog. Theor. Phys. {\bf 104} 867, (2000), hep-th/0005197; 
``{\em   Scheme dependence of holographic conformal anomaly in 5-d gauged supergravity with nontrivial bulk potential,}''
Phys. Lett. {\bf B494} 318, (2000), hep-th/0009015.
\bibitem{marteli2}
D. Martelli and A. Miemiec,
``{\em CFT / CFT interpolating RG flows and the holographic c function,}''
JHEP {\bf 0204} 027, (2002), hep-th/0112150.

\bibitem{berg}
M. Berg and H. Samtleben,
``{\em An Exact holographic RG flow between 2-d conformal fixed points,}''
JHEP {\bf  0205} 006, (2002), hep-th/0112154.
\bibitem{Zamo} A. Zamolodchikov, 
``{\it Irreversibility of the Flux of the Renormalization Group in a 2D 
Field Theory}", 
JETP Lett. {\bf 43} (1986) 730.
\bibitem{Ostrogradsky} 
    M. V. Ostrogradsky, Mem. Acad. Sci.,
    ``{\it M{\' e}moire sur les {\' e}quations diff{\' e}rentielles relatives 
    au probl{\` e}me des  isoperim{\` e}tre}", 
	St. Petersberg {\bf 6} (1850) 385.

\bibitem{BL} I. L. Buchbinder and S. L. Lyahovich, 
    `` {\it Canonical quantization and local measure of $R^{2}$ gravity}", 
	Class. Quantum Grav. {\bf 4} (1987) 1487.

\bibitem{BLK} I. L. Buchbinder, S. L. Lyahovich, and V. A. Krykhtin,
``{\it Canonical quantization of topologically massive gravity}",
	Class. Quantum Grav. {\bf 10} (1993) 2083.
\bibitem{chiral-c}  F. Bastianelli and
	U. Lindstr${\rm \ddot{o}}$m, ``{\em C-theorem for two dimensional
	chiral theories},'' arXiv: hep-th/9604001.
\bibitem{AW} L. Alvarez-Gaume and E. Witten, ``{\em Gravitational
	Anomalies}'', Nucl. Phys. {\bf B234} (1984) 269.
\bibitem{BZ} W. A. Bardeen and B. Zumino, ``{\em  Consistent and
	Covariant Anomalies in Gauge and Gravitational Theories}'',
	Nucl. Phys. {\bf B244} (1984) 421.
\bibitem{Bertlmann}
R. A. Bertlmann, ``{\em Anomalies in Quantum Field Theory}'', Oxford
	university press (1996).
\bibitem{Graham-Lee} C. R. Graham and J. M. Lee, ``{\em Einstein Metrics
	with Prescribed Conformal Infinity on the Ball,}''
	Adv. Math. {\bf 87} (1991), 186.
\end{thebibliography}
\end{document}